\documentclass[aps,pre,preprint,groupedaddress,showpacs]{revtex4}

\usepackage{amsmath}
\usepackage{amssymb}
\usepackage{epsfig}
\newcommand {\dr}{{\mathrm d}\mathbf{r}}

\newcommand {\rr}{\mathbf{r}}

\newcommand {\Tr}{{\mathrm T}{\mathrm r}}

\newcommand {\q}{0}

\begin{document}

\title{Soft core fluid in a quenched matrix of soft core particles:\\
       A mobile mixture in a model gel}

\author{A.J. Archer$^1$}
\email[]{Andrew.Archer@bristol.ac.uk}
\author{M. Schmidt$^{1,2}$} 
\author{R. Evans$^1$}
\affiliation{1. H.H. Wills Physics Laboratory, University of Bristol,
Tyndall Avenue, Bristol BS8 1TL, UK\\
2. Institut f{\"u}r Theoretische Physik II,
  Heinrich-Heine-Universit{\"a}t D{\"u}sseldorf,
  Universit{\"a}tsstra{\ss}e 1, D-40225 D{\"u}sseldorf, Germany.}

\date{\today}

\begin{abstract}
  We present a density--functional study of a binary phase--separating
  mixture of soft core particles immersed in a random matrix of
  quenched soft core particles of larger size. This is a model for a
  binary polymer mixture immersed in a crosslinked rigid polymer
  network. Using the replica `trick' for quenched--annealed mixtures we
  derive an explicit density functional theory that treats the
  quenched species on the level of its one-body density distribution.
  The relation to a set of effective external potentials acting on the
  annealed components is discussed.  We relate matrix--induced
  condensation in bulk to the behaviour of the mixture around a single
  large particle. The interfacial properties of the binary mixture at
  a surface of the quenched matrix display a rich interplay between
  capillary condensation inside the bulk matrix and wetting phenomena
  at the matrix surface.
\end{abstract}

\pacs{78.55.Mb, 61.20.Gy, 64.10.+h}

\maketitle

\section{Introduction}
\label{sec:intro}
Spatial confinement can have a dramatic effect on the properties of a
fluid. The combined effects of finite pore size and substrate fluid interactions
can alter the location of phase boundaries from what is found for the bulk
adsorbate. Many studies have focused on the equilibrium structure and phase
behaviour of fluids confined in model pores or capillaries and phenomena such as
capillary condensation (the shifted gas--liquid transition) and capillary
freezing are rather well--understood for a one--component fluid confined in a
single, infinitely long slit or cylinder \cite{evans90,gelb99}. The kinetics of
nucleation accompanying capillary condensation has also been investigated
\cite{talanquer00}. When the adsorbate consists of more than one species,
capillary induced fluid--fluid demixing can occur in addition to capillary
condensation -- see e.g.\ \cite{woywod03pre} for recent work.
However, most porous media exhibit complicated, interconnected pore shapes
\cite{gelb99} and cannot
be modelled as idealised capillaries. Rather, a random porous
network structure provides a more realistic starting model.
In recent years the direct treatment of fluids adsorbed
random pore structures has been achieved within the framework of lattice fluids
yielding significant new insight into adsorption phenomena
\cite{kierlik01,kierlik02,sarkisov02,detcheverry03,woo03,schmidt03lfmf}.

An alternative way of modelling porous structures
is to consider pore structures generated by quenched
configurations of model fluids. These configurations carry a
statistical weight that is determined solely by the Hamiltonian of the
(pure) model fluid. The adsorbate fluid is then brought
into contact with the quenched particles (referred to as the matrix),
and is ``annealed'' in the sense that it is under the influence
of the matrix as an external potential. In addition to the ensemble
average, an average over all realisations of the matrix must be
carried out. Frequently such models are tackled using liquid state integral
equation theory, relying on the replica Ornstein--Zernike equations
\cite{maddenglandt,givenstell} -- an approach that works on the level
of two--body correlation functions.

Recently a classical density functional theory
(DFT) \cite{evans79,evans92}, operating on the one--body level, was
shown to be able to treat consistently quenched--annealed
mixtures. This quenched--annealed DFT (or replica DFT)
\cite{schmidt02pordf,reich04poroned} treats both the annealed and the
quenched species on the level of their one--body density
distributions. Hence the complicated external potential that the
quenched particles exert on the fluid never enters explicitly into the
theoretical framework. This approach
constitutes an enormous simplification as far as
practical computations are concerned. Investigations were carried out
for hard core models and for the quenched--annealed
Asakura--Oosawa--Vrij model of
colloid--polymer mixtures \cite{wessels05matrixint}.

The Gaussian core model (GCM) \cite{likos01II} belongs to a different
class of system, characterised by fully penetrable pair interactions,
i.e.\ at zero spatial separation between the centres of a pair of
particles the potential energy is finite and is comparable in magnitude
to the thermal energy. The fact that a very simple density
functional approximation is accurate at high fluid densities
\cite{likos01II,lang00,louis00II,archer01,archer04,patrykiejew04} has
facilitated a number of detailed investigations and much is known about
the bulk and interfacial properties as well as subtle solvation phenomena, in
this model fluid \cite{archer01,archer02,archer02el,archer03,archer04,archer05}.

In the present work, we are concerned with a quenched--annealed GCM
modelling a binary polymer solution as an adsorbate in a quenched
(crosslinked) rigid polymer network -- i.e.\ a mobile binary mixture
adsorbed in a model gel.  Quenching components of a GCM mixture
results in {\em soft confinement} of the annealed species, as opposed
to the hard confinement considered with replica DFT in previous studies.
The particularly simple (RPA--like) form of the excess free energy
functional enables us to elucidate the structure of the replica DFT.
We can relate the effect of the quenched component directly to
the influence of an effective external potential acting on the fluid
particles; this is shown to take the form
of a convolution of the matrix
one--body density profile and the matrix--fluid (pair) interaction
potential.

The paper is organised as follows. In Sec.\ \ref{sec:QA} we give a brief
overview of the statistical mechanics of quenched--annealed mixtures.
In Sec.\ \ref{sec:GCM} the DFT for the quenched--annealed GCM is
presented.  The case of matrices with uniform density is covered in
Sec.\ \ref{sec:uniform}. We present results for bulk properties of a
particular mixture in Sec.\ \ref{sec:particular_mixture} and treat
inhomogeneities induced by the surface of a matrix in Sec.\
\ref{sec:matrix_surface}. Discussion and conclusions are presented in
Sec.\ \ref{sec:conclusions}.

\section{Quenched--annealed systems}
\label{sec:QA}

We consider a system composed of three different species of
particles. One of the species, labelled $\q$, is quenched, forming a
matrix into which the remaining two species, labelled 1 and 2, are
annealed, i.e.\ the particles of species $\q$ form an immobile
network in which the other species are free to move.  The correlations
between the quenched matrix particles are independent of the adsorbate
particles and are determined solely by interactions between matrix
particles.  We denote particle numbers by $N_a$, where $a=\q,1,2$
labels the different species, and denote the sets of position
coordinates of particles of species $a$ by $\rr^{N_a} \equiv \{
\rr_{i,a}, i=1,\ldots,N_a \}$. The total external potential energy of
species $a$, the total potential due to pair interactions between
particles of species $a$, and the total potential due to pair
interactions between particles of (unlike) species $a$ and $b$, are
defined, respectively, by
\begin{eqnarray}
V_a(\rr^{N_a})&=&\sum_{i=1}^{N_a}V_a^{\rm ext}(\rr_i),
\label{eq:V_a} \\
\Phi_{aa}(\rr^{N_a})
&=&\sum_{i=1}^{N_a}\sum_{j=i+1}^{N_a}v_{aa}(|\rr_i-\rr_j|),
\label{eq:phi_aa} \\
\Phi_{ab}(\rr^{N_a},\rr^{N_b})
&=&\sum_{i=1}^{N_a}\sum_{j=1}^{N_b}v_{ab}(|\rr_i-\rr_j|), \quad a \neq b
\label{eq:phi_ab}
\end{eqnarray}
where $V_a^{\rm ext}(\rr_i)$ is the external potential acting on particle
$i$ of species $a$ and $v_{ab}(|\rr_i-\rr_j|)$ is the pair potential
between particle $i$ of species $a$ and particle $j$ of species $b$.
It is important to note that for quenched--annealed systems the
potentials $\Phi_{\q 1}$ and $\Phi_{\q 2}$, exerted by the quenched
particles on the annealed matrix particles, never enter into the
description of any properties of the matrix (i.e.\ species 0) alone;
the latter originate solely from $V_\q(\rr^{N_0})$ and
$\Phi_{\q\q}(\rr^{N_0})$.  

Hence, the partition function for the quenched matrix particles is
\begin{equation}
\Xi_\q(\mu_\q)=\Tr_{N_\q}\exp[-\beta(\Phi_{\q\q}+V_\q)]
\label{eq:Xi_q}
\end{equation}
where $\mu_\q$ is the chemical potential for the quenched particles
and the classical trace is
\begin{equation}
\Tr_{N_a}(\cdot)
=\sum_{N_a=0}^{\infty} \frac{z_a^{N_a}}{N_a!} \int \dr_a^{N_a}(\cdot),
\label{eq:Tr}
\end{equation}
where $z_a=\Lambda_a^{-3} \exp( \beta \mu_a)$ is the fugacity and
$\Lambda_a$ is the thermal de Broglie wavelength for species $a$, and
$\beta=1/k_BT$, where $k_B$ is the Boltzmann constant. Strictly,
$\Xi_\q=\Xi_\q(\mu_\q,T,V)$ is a
function of the chemical potential $\mu_\q$, the temperature, $T$, and
the volume of the system, $V$. However, we
will suppress the dependence on $T$ and $V$ for notational
convenience. The grand potential for the quenched particles is then
given by
\begin{equation}
\Omega_\q(\mu_\q)=-k_BT \ln \Xi_\q(\mu_\q).
\label{eq:Omega_q}
\end{equation}
For a {\em given} configuration of matrix particles, $\rr^{N_\q}$, the
grand potential for the annealed particles, adsorbed in the matrix, is:
\begin{equation}
\Xi_{12}(\rr^{N_\q},\mu_1,\mu_2)=\Tr_{N_1}\Tr_{N_2}
\exp[-\beta(V_1+V_2+\Phi_{12}+\Phi_{\q1}+\Phi_{\q2})],
\label{eq:Xi_12}
\end{equation}
and the grand potential for this configuration is
\begin{equation}
\tilde{\Omega}_{12}(\rr^{N_\q},\mu_1,\mu_2)
=-k_BT \ln \Xi_{12}(\rr^{N_\q},\mu_1,\mu_2).
\label{eq:Omega_12_tilde}
\end{equation}
We now average this grand potential over all configurations of
quenched matrix particles:
\begin{equation}
\Omega_{12}(\mu_\q,\mu_1,\mu_2)=\Xi_\q^{-1} \Tr_{N_\q}
\tilde{\Omega}_{12}(\rr^{N_\q},\mu_1,\mu_2)\exp[-\beta (\Phi_{00}+V_0)].
\label{eq:Omega_12}
\end{equation}
$\Omega_{12}$ is a key quantity: it determines many of the properties of
the fluid in the matrix. We now apply the replica `trick' that proves to
be a powerful idea to actually calculate $\Omega_{12}$
\cite{schmidt02pordf,schmidt02aom,reich04poroned}.

The idea is to introduce $s$ `replicas', or copies, of the sets of
particles of species 1 and 2 into the matrix of quenched particles.
Particles do not interact with particles of a different replica, but
only with particles of the same replica as well as with the matrix
particles. The partition sum for this $(2s+1)$--component mixture is:
\begin{equation}
\Xi(\mu_\q,\mu_1,\mu_2;s)=\Tr_{N_\q} \exp[-\beta(\Phi_{\q\q}+V_\q)]
\left\{ \Tr_{N_1}\Tr_{N_2}
\exp[-\beta(V_1+V_2+\Phi_{12}+\Phi_{\q1}+\Phi_{\q2})] \right\}^s
\label{eq:Xi}
\end{equation}
and the corresponding grand potential follows as
\begin{equation}
\Omega(\mu_\q,\mu_1,\mu_2;s)=-k_BT \ln \Xi(\mu_\q,\mu_1,\mu_2;s).
\label{eq:Omega}
\end{equation}
Performing an analytic continuation in $s$, and noting that $\lim_{s
\rightarrow 0} \mathrm{d} x^s/\mathrm{d} s=\ln x$, we find
\cite{reich04poroned}
\begin{equation}
\Omega_{12}(\mu_\q,\mu_1,\mu_2)=\lim_{s \rightarrow 0}
\frac{\partial}{\partial s} \Omega(\mu_\q,\mu_1,\mu_2;s).
\label{eq:Omega_seq0}
\end{equation}
Thus we have constructed a means of tackling quenched--annealed systems
by considering the
corresponding replicated (and fully annealed) mixture. At this point
we can make connection with DFT, which provides a powerful approach
to calculating the grand potential for inhomogeneous systems.  In DFT
the grand potential is expressed as a functional of the fluid one--body
density profiles, $\rho_a(\rr)$
\cite{evans79,evans92}. The grand potential in
(\ref{eq:Omega}) is obtained by considering a {\em variational} grand
potential functional, $\Omega^{\rm var}([ \{\rho_a \} ],\{\mu_a\};s)$,
which is minimised with respect to the density profiles
\cite{evans79,evans92},
\begin{equation}
\frac{\delta \Omega^{\rm var}([ \{\rho_a \} ],\{\mu_a\};s)}
{\delta \rho_a(\rr)}=0.
\label{eq:min_cond}
\end{equation}
Inserting the density profiles that are the solutions to
(\ref{eq:min_cond}) into $\Omega^{\rm var}$ yields as the minimal
value the true grand potential,
$\Omega(\mu_\q,\mu_1,\mu_2;s)$. Performing an analytic continuation in
$s$, and Taylor expanding $\Omega^{\rm var}$ around $s=0$, gives
\begin{equation}
\Omega^{\rm var}([ \{\rho_a \} ],\{\mu_a\};s) = 
\Omega_\q^{\rm var}( [\rho_\q],\mu_\q)
+s \Omega_{12}^{\rm var}([\{\rho_a \} ],\{\mu_a\}) + {\cal O}(s^2).
\label{eq:taylor}
\end{equation}
$\Omega_\q^{\rm var}( [\rho_\q],\mu_\q)$ is simply the grand potential
for the pure system of matrix particles; furthermore $\Omega_{12}^{\rm
var}([\{\rho_a \} ],\{\mu_a\})=\lim_{s \rightarrow 0} \partial
\Omega^{\rm var}([\{\rho_a \} ],\{\mu_a\};s)/ \partial s$.  The key
point is that $\Omega_{12}^{\rm var}$ in (\ref{eq:taylor}) corresponds
to an average
over the different matrix configurations and is hence only a
functional of the matrix density profile $\rho_\q(\rr)$, rather than
depending on particular matrix configurations $\rr^{N_\q}$.
Explicitly, we can write the functionals on the right hand side of
(\ref{eq:taylor}) as
\begin{eqnarray}
\Omega_\q^{\rm var}([\rho_\q(\rr)],\mu_\q)={\cal F}_\q^{\rm id}[\rho_\q(\rr)]
+{\cal F}_\q^{\rm ex}[\rho_\q(\rr)]
-\int \dr \rho_\q(\rr)[\mu_\q-V_\q^{\rm ext}(\rr)],
\label{eq:grand_q_var}
\end{eqnarray}
where
\begin{eqnarray}
{\cal F}_a^{\rm id}[\rho_a(\rr)]
=k_BT \int \dr \rho_a(\rr)[\ln(\Lambda_a^3 \rho_a(\rr))-1]
\label{eq:F_id}
\end{eqnarray}
is the Helmholtz free energy of the ideal gas and ${\cal F}_\q^{\rm
ex}[\rho_\q(\rr)]$ is the excess contribution to the total Helmholtz
free energy arising from interactions between the matrix
particles. Similarly,
\begin{eqnarray}
\Omega_{12}^{\rm var}([\rho_\q,\rho_1,\rho_2],\mu_1,\mu_2)=
{\cal F}_{1}^{\rm id}[\rho_1(\rr)]+{\cal F}_{2}^{\rm id}[\rho_2(\rr)]
+{\cal F}_{12}^{\rm ex}[\rho_\q(\rr),\rho_1(\rr),\rho_2(\rr)] \notag \\
-\sum_{a=1}^2 \int \dr \rho_a(\rr)[\mu_a-V_a^{\rm ex}(\rr)],
\label{eq:grand_quench_1}
\end{eqnarray}
where ${\cal F}_a^{\rm id}[\rho_a(\rr)]$, $a=1,2$ is given by Eq.\
(\ref{eq:F_id}) and ${\cal F}_{12}^{\rm ex}[\rho_\q,\rho_1,\rho_2]$
is the contribution to the Helmholtz free energy from interactions
between the particles. This quantity {\em includes} interactions between
the annealed particles and the quenched matrix particles. In general,
${\cal F}^{\rm ex}_{12}$ is an unknown functional. However, depending
on the form of the interactions between the particles, there exist a
number of accurate approximate functionals \cite{evans92}.

The variational principle (\ref{eq:min_cond}), when applied to (\ref{eq:taylor}),
becomes \cite{schmidt02pordf,reich04poroned}:
\begin{eqnarray}
\frac{\delta \Omega_\q^{\rm var}([ \rho_\q ],\mu_\q)}{\delta \rho_\q(\rr)}&=&0,
\label{eq:min_cond_2a} \\
\frac{\delta \Omega_{12}^{\rm var}([\rho_\q,\rho_1,\rho_2],\mu_1,\mu_2)}
     {\delta \rho_a(\rr)}
 &=&0, \quad a=1,2.
\label{eq:min_cond_2b}
\end{eqnarray}
One therefore uses (\ref{eq:min_cond_2a}), together with
(\ref{eq:grand_q_var}) and a suitable approximation for ${\cal
F}_\q^{\rm ex}[\rho_\q(\rr)]$, to first calculate the average one body
density for the quenched matrix particles, $\rho_\q(\rr)$. Then one
uses this density profile as input to calculate the density profiles
$\rho_1(\rr)$ and $\rho_2(\rr)$ of the annealed species using Eqs.\
(\ref{eq:grand_quench_1}) and (\ref{eq:min_cond_2b}), again with a
suitable approximation for ${\cal F}_{12}^{\rm
ex}[\{\rho_a(\rr)\}]$. In the next section we shall describe specific
approximations for particles that interact via repulsive Gaussian pair
potentials.

\section{DFT for the quenched--annealed GCM}
\label{sec:GCM}

The interactions between macromolecules such as
polymers or dendrimers in
solution can be modelled by means of an effective pair potential
between the centres of mass of the macromolecules. This implies
treating each macromolecule as a (soft) `particle'.  For a
good solvent the effective pair potential is purely repulsive
and is well--approximated by a repulsive Gaussian form
\cite{likos01II,dautenhahn94,louis00,bolhuis01,
louis02physica,bolhuis02macromolecules,likos02jcp,likos01dendrimers}. In
the GCM, the potential between particles of species $a$ and $b$
is given by:
\begin{equation}
v_{ab}(r)=\epsilon_{ab}\exp(-r^2/R_{ab}^2)
\label{eq:pairpot}
\end{equation}
where $\epsilon_{ab}$ is the energy penalty (due to entropic effects on
the segment level) for complete overlap of the centres of mass.
Typically $\epsilon_{ab} \sim
2k_BT$ and $R_{aa} \sim R_{g,a}$, the radius of gyration of particles
of the $a^{th}$ species. The pair potential parameters for the
potentials between unlike--species, as obtained from simulations of
model polymer solutions \cite{dautenhahn94}, are consistent with the
(non--additive) mixing rules,
\begin{equation}
R_{12}^2=\frac{1}{2}(R_{11}^2+R_{22}^2)
\label{eq:pair_pot_mix_rule_1}
\end{equation}
and
\begin{equation}
\epsilon_{12}\lesssim \epsilon_{11}=\epsilon_{22}.
\label{eq:pair_pot_mix_rule_2}
\end{equation}

For the GCM the mean--field excess Helmholtz free energy functional,
\begin{equation}
{\cal F}^{\rm ex}[\{ \rho_a(\rr) \}]=\frac{1}{2}\sum_{a,b} \int \dr \int
\dr' \rho_a(\rr) \rho_b(\rr')v_{ab}(|\rr-\rr'|),
\label{eq:F_ex}
\end{equation}
is known to be accurate at high densities
\cite{likos01II,lang00,louis00II,archer01,archer04,patrykiejew04}
when each particle interacts with a large number
of neighbouring particles -- the classic mean field situation. The
functional (\ref{eq:F_ex}) generates the RPA closure for the pair
direct correlation functions upon differentiation
\cite{likos01II,evans92} since
\begin{equation}
c^{(2)}_{ab}(\rr,\rr') \equiv
-\beta \frac{\delta^2 {\cal F}^{\rm ex}[\{ \rho_a(\rr) \}]}
{\delta \rho_a(\rr) \delta \rho_b(\rr')}
=-\beta v_{ab}(|\rr-\rr'|).
\label{eq:c2}
\end{equation}
Here we focus on the case of a ternary mixture where one
species of GCM particles is quenched (treating cases with larger
numbers of both annealed and quenched components requires
straightforward generalisations). The quenched particles have a
one--body density distribution, $\rho_\q(\rr)$, which is obtained by
minimising (\ref{eq:grand_q_var}) using the one component ($\q$)
version of (\ref{eq:F_ex}). The density distributions of the remaining
(annealed) species of particles are obtained by minimising the grand
potential functional:
\begin{eqnarray}
\Omega_{12}^{\rm var}[\{\rho_a(\rr)\}]&=&{\cal F}_1^{\rm id}[\rho_1(\rr)]
+{\cal F}_2^{\rm id}[\rho_2(\rr)] \notag \\&&
+\frac{1}{2}\sum_{n,m=1}^2 \int \dr \int \dr'
\rho_n(\rr) \rho_m(\rr')v_{nm}(|\rr-\rr'|) \notag \\&&
+\sum_{n=1}^2 \int \dr \int \dr'
\rho_n(\rr) \rho_\q(\rr')v_{\q n}(|\rr-\rr'|) \notag \\&&
-\sum_{n=1}^2 \int \dr \rho_n(\rr)[\mu_n-V_n^{\rm ext}(\rr)].
\label{eq:grand_quench_2}
\end{eqnarray}
Expressing the density distribution of the quenched particles as
$\rho_\q(\rr)=\rho_\q^b+\Delta \rho_\q(\rr)$, i.e.\ as a modulation
$\Delta \rho_\q(\rr)$ around a constant density $\rho_\q^b$,
(\ref{eq:grand_quench_2}) can be written as:
\begin{eqnarray}
\Omega_{12}^{\rm var}[\{\rho_a(\rr)\}]={\cal F}_1^{\rm id}[\rho_1(\rr)]
+{\cal F}_2^{\rm id}[\rho_2(\rr)]
+{\cal F}^{\rm ex}[\{\rho_n(\rr)\}] \notag \\
-\sum_{n=1}^2 \int \dr \rho_n(\rr)[\mu_n^{\rm eff}-V_n^{\rm eff}(\rr)]
\label{eq:grand_2}
\end{eqnarray}
where the effective chemical potentials are
\begin{eqnarray}
\mu_n^{\rm eff}=\mu_n-\rho_\q^b \hat{v}_{\q n}(0),
\label{eq:mu_eff}
\end{eqnarray}
and $\hat{v}_{\q n}(0) \equiv \int \dr v_{\q
n}(r)=\pi^{3/2}\epsilon_{nq} R_{nq}^3$ is the $k \rightarrow 0$ limit
of the Fourier transform of the pair potential $v_{\q n}(r)$. The
effective external potentials are
\begin{eqnarray}
V_n^{\rm eff}(\rr)=V_n^{\rm ext}(\rr)+\int \dr'
\Delta \rho_\q(\rr')v_{\q n}(|\rr-\rr'|).
\label{eq:V_eff}
\end{eqnarray}
This analysis demonstrates that the quenched matrix has a very similar
effect to that of an external potential acting on the mobile
components. However, because of the convolution form in Eq.\
(\ref{eq:V_eff}), the matrix effective potentials $V_n^{\rm eff}(\rr)$
are quite different from the external potentials that are typically
tackled within DFT, such as the hard potentials modelling container
walls.

\section{A matrix of uniform density}
\label{sec:uniform}

As a prerequisite for our subsequent interface study we consider the
case where the matrix density profile is constant:
$\rho_\q(\rr)=\rho_\q^b$ (i.e.\ $\Delta \rho_\q(\rr)=0$). In the
particular case where the
external potentials also vanish, the one body densities of the
annealed species are constants: $\rho_n(\rr)=\rho_n^b$, $n=1,2$. As
convenient variables we use the total (bulk) density of the annealed
species, $\rho^b=\rho_1^b+\rho_2^b$, and the relative concentration,
$x \equiv \rho_2^b/\rho^b$.  From Eq.\ (\ref{eq:grand_quench_2}) the
Helmholtz free energy per particle, $f_{12}=F/N$, for the annealed GCM
fluid is simply \cite{archer01}:
\begin{eqnarray}
f_{12}(\rho^b,x)=f_{12}^{\rm id}(\rho^b,x)+\frac{1}{2} \rho^b \hat{V}_0(x)
+ \rho_\q^b Q(x),
\label{eq:f}
\end{eqnarray}
where $f_{12}^{\rm id}$ is the ideal gas contribution,
\begin{eqnarray}
f_{12}^{\rm id}(\rho^b,x)&=&(1-x)\ln(1-x)+x \ln x \notag \\
&& +(1-x) \ln(\Lambda_1^3 \rho^b) +x \ln(\Lambda_2^3 \rho^b)-1,
\label{eq:f_id}
\end{eqnarray}
the interaction term is
\begin{eqnarray}
\hat{V}_0(x)=(1-x)^2 \hat{v}_{11}(0)+2x(1-x)\hat{v}_{12}(0)+x^2 \hat{v}_{22}(0),
\label{eq:Vhat_0}
\end{eqnarray}
and the contribution due to the matrix is
\begin{eqnarray}
Q(x)=(1-x)\hat{v}_{1\q}(0)+x\hat{v}_{2\q}(0).
\label{eq:Q_of_x}
\end{eqnarray}
The latter has a particularly simple form; it is independent
of $\rho^b$ and only linearly dependent on $x$.  This leads to several
special features of the fluid inside the matrix. First, the
pressure of the annealed GCM fluid in the matrix is independent of the
matrix density $\rho_\q^b$. As the pressure $P$ is obtained from
\begin{eqnarray}
P=-\left( \frac{\partial f_{12}}{\partial v} \right)_{x,T},
\label{eq:pressure}
\end{eqnarray}
where $v=1/\rho^b$ is the volume per particle and the matrix
contribution to the free energy (\ref{eq:f}) is independent of
$\rho^b$, it follows that $P$ is independent of $\rho_\q^b$.

Second, the phase boundaries in the $(x,\rho^b)$
plane of the fluid inside the matrix are the same as those without the
matrix (it is known that the binary GCM fluid can exhibit fluid--fluid
phase separation \cite{louis00II,archer01}).
This can be seen as follows: the locus of the spinodal line
can be obtained from the Gibbs free energy per particle,
$g_{12}=f_{12}+Pv$, via the condition:
\begin{eqnarray}
\left( \frac{\partial^2 g_{12}}{\partial x^2} \right)_{P,T}=0.
\label{eq:spin_cond}
\end{eqnarray}
Since $P$ is independent of $\rho_\q^b$ and the terms in $f_{12}$ (and
therefore in $g_{12}$) that are functions of $\rho_\q^b$ are only
linearly dependent on $x$, we see from (\ref{eq:spin_cond}) that in
the $(x,\rho^b)$ plane the spinodal line will be the same for all
values of $\rho_\q^b$. The binodal (coexistence curve) is obtained
from the conditions of equal pressure $P$ and equal chemical potentials
in the coexisting phases,
\begin{eqnarray}
P_{\alpha}=P_{\beta} \notag \\
\mu_{1,\alpha}=\mu_{1,\beta} \notag \\
\mu_{2,\alpha}=\mu_{2,\beta},
\label{eq:bin_cond}
\end{eqnarray}
where $\alpha$ and $\beta$ label the coexisting phases. From Eq.\
(\ref{eq:mu_eff}), we see that the presence of the matrix results in only
a linear shift in the chemical potentials of the fluid particles, and
this is the
same in both fluid phases, $\alpha$ and $\beta$. Therefore, the
binodal curve for the binary fluid in the matrix is at the same
location in the $(x,\rho^b)$ plane as it is without the matrix.

The special form of the free energy
makes for a simple analysis of the situation when one
has a bulk binary GCM fluid acting as a reservoir and coupled to the
system inside the matrix that consists
of quenched particles of a third species of
GCM particles. The chemical potentials, $\mu_1$ and $\mu_2$, are set
by the fluid in the bulk reservoir. One can calculate $\rho_1^b$ and
$\rho_2^b$, the densities of
the fluid in the matrix, using Eq.\ (\ref{eq:mu_eff}), i.e.\ the
densities of the fluid in the matrix are simply those of a bulk fluid
with chemical potentials equal to $\mu_n^{\rm eff}$, $n=1,2$, given by
Eq.\ (\ref{eq:mu_eff}).

For a binary GCM fluid exhibiting phase separation, this means that
the chemical potentials, $\mu_n$, can correspond to a fluid on one side
of the binodal of the reservoir (say, the $\alpha$--phase), but the
effective chemical potentials $\mu_n^{\rm eff}$, given by Eq.\
(\ref{eq:mu_eff}), for the fluid inside the matrix can correspond to
the other side of the binodal (the $\beta$--phase). Under these
circumstances
a (capillary) condensation of the coexisting $\beta$--phase occurs in the
matrix. Such condensation transitions are well--known for fluids in
porous media (the matrix)
\cite{evans90,gelb99,kierlik01,kierlik02,sarkisov02,detcheverry03}.
Perhaps the best known case is that of a gas condensing to a liquid at
pressures below the bulk saturated vapour pressure or, for fixed
pressure, at temperatures above the bulk boiling temperature.
In this case the bulk $\alpha$--phase corresponds to the gas and the
$\beta$--phase to the liquid \cite{vycor}.

Owing to the simplicity of the present model fluid and the theoretical
description
that we use, Eq.\ (\ref{eq:mu_eff}) provides a very simple mapping
between points in the $(x,\rho^b)$ plane corresponding to the density
and concentration of the fluid in the bulk reservoir and the same
quantities for the fluid in the matrix. For a given matrix density (and
given parameters of the pair potential between matrix and fluid
particles) we can easily determine the matrix condensation phase
behaviour.

\section{Results for a particular binary mixture}
\label{sec:particular_mixture}

\begin{figure}
\includegraphics[width=8cm]{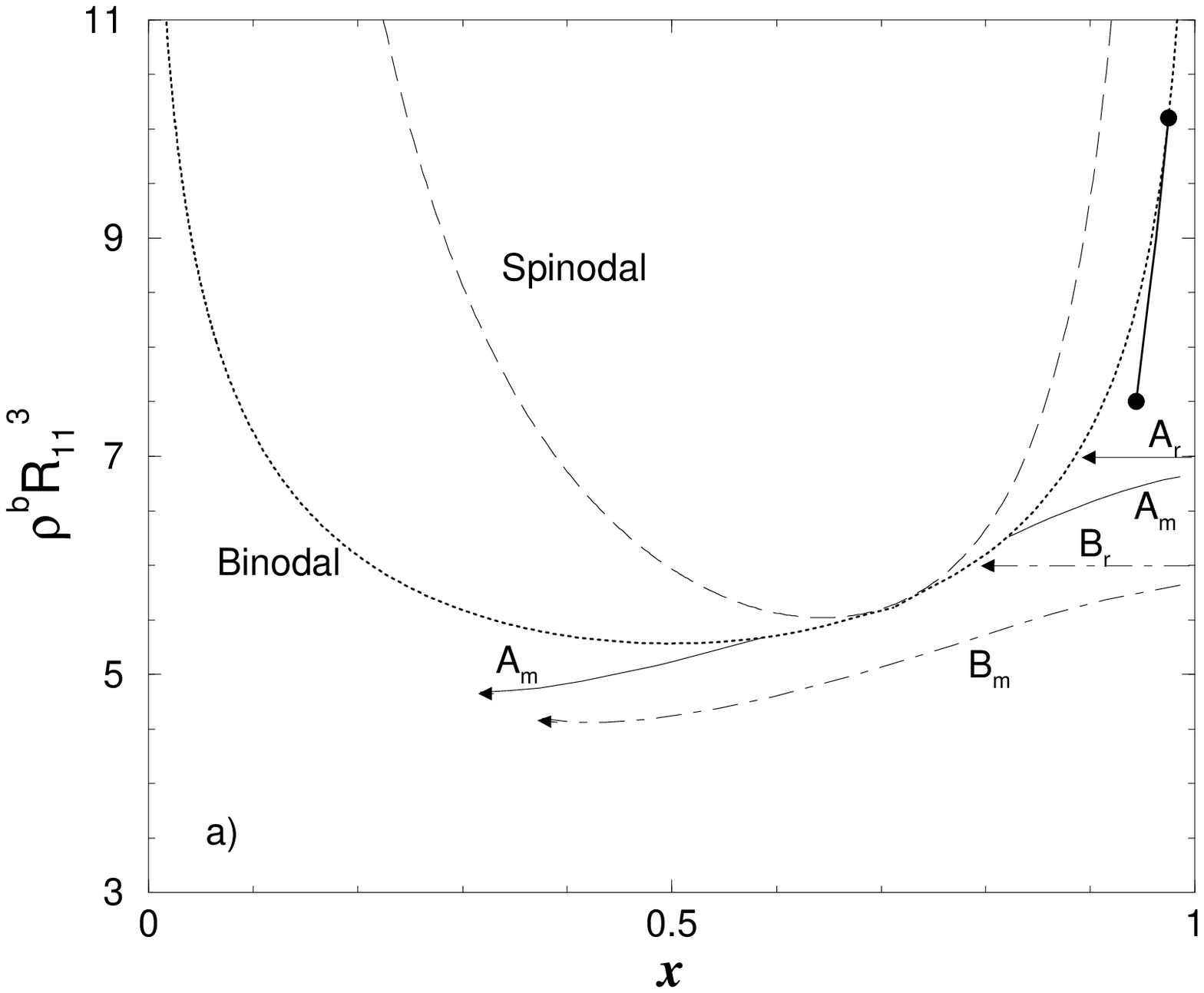}
\includegraphics[width=8cm]{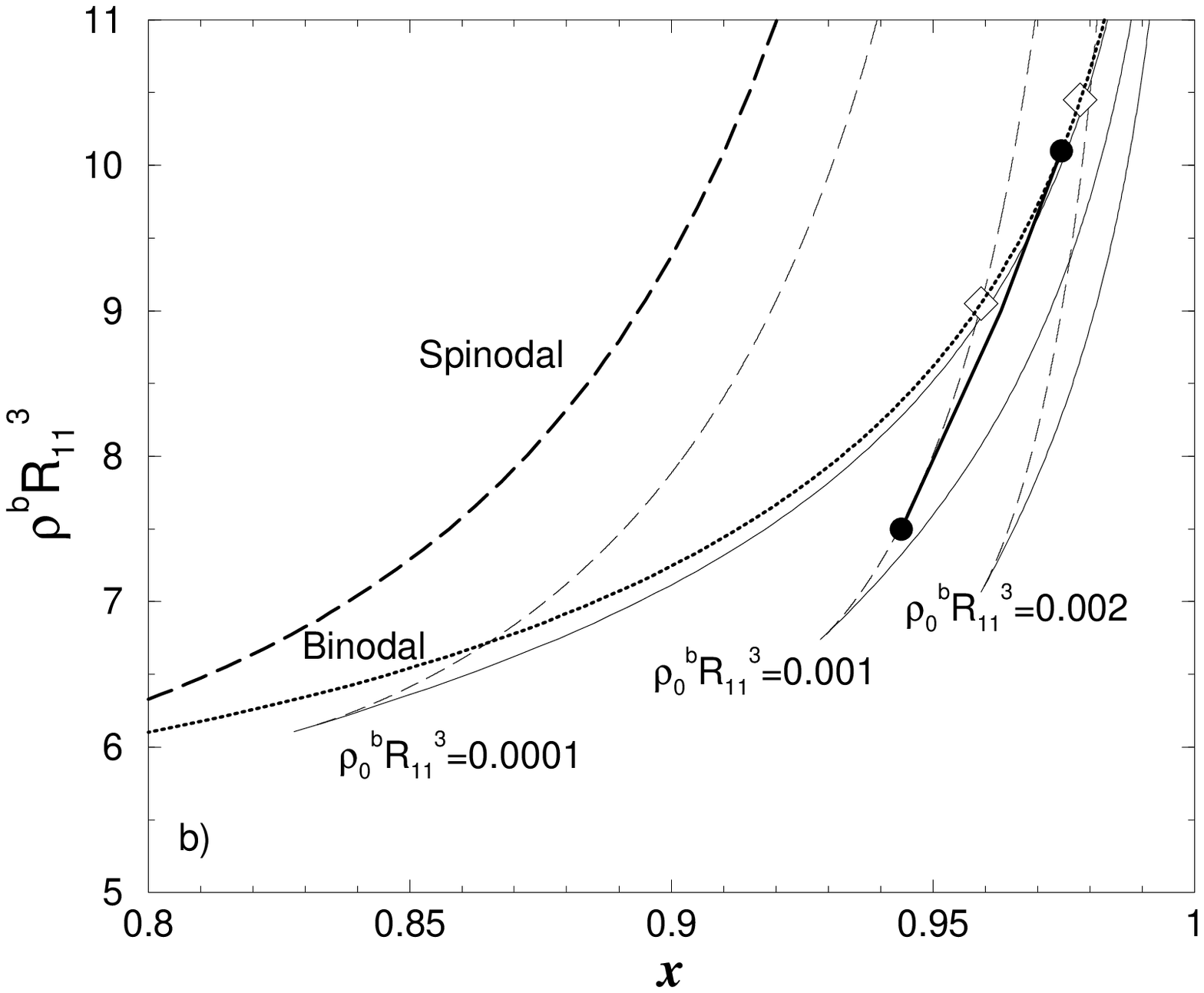}
\caption{The bulk phase diagram for a binary mixture of GCM particles
with $\epsilon_{12}/ \epsilon_{11} =0.944$ and $R_{22}/R_{11}=0.665$
(see also Ref.\ \cite{archer01}) in terms of the total density,
$\rho^b$, and the concentration, $x$, of the smaller species 2. In a)
the paths ${\mathrm A_r}$ and ${\mathrm B_r}$ (arrows) are at constant
total density in the
reservoir. The paths ${\mathrm A_m}$ and ${\mathrm B_m}$ 
indicate the corresponding paths for the fluid in a
matrix of density $\rho_\q^bR_{11}^3=10^{-3}$.  b) shows an expanded
view of the phase diagram. The three additional pairs of binodal (solid
line) and spinodal (dashed) lines show the loci in the {\em reservoir}
phase diagram which
correspond [via Eq.\ (\ref{eq:mu_eff})] to the binodal/spinodal of
the fluid adsorbed in the matrix, plotted for three different matrix
densities. In a) and b) the solid line whose ends are denoted by filled
circles is
the thin--thick adsorbed film transition of the binary fluid adsorbed
around a {\em single} big GCM particle with pair potential parameters
$\epsilon_{\q1}=k_BT$, $\epsilon_{\q2}=0.8k_BT$, $R_{\q1}/R_{11}=5.0$ and
$R_{\q2}/R_{11}=4.97$ -- see Refs.\ \cite{archer03,archer05}. The two
points ($\diamond$) are where the spinodal for this single
particle thin--thick adsorbed transition meets the bulk fluid binodal.}
\label{fig:1}
\end{figure}

\begin{figure}
\includegraphics[width=8cm]{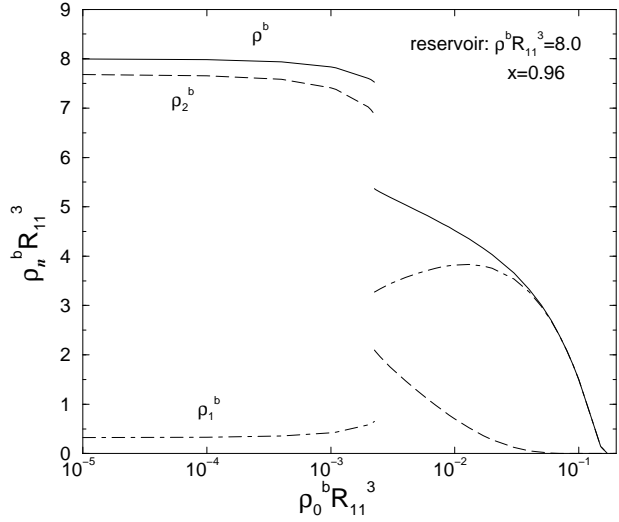}
\caption{Fluid densities, $\rho_n^b$, inside the matrix
corresponding to a reservoir with fixed total
density $\rho^b R_{11}^3=8.0$ and
concentration $x=0.96$ plotted versus (uniform) matrix density
$\rho_\q^b$. For $\rho_\q^b R_{11}^3>2.3\times
10^{-3}$, condensation of the coexisting
fluid phase occurs in the matrix. Note that $\rho_1^b$ is a
non--monotonic function of $\rho_\q^b$, whereas $\rho_2^b$
and the total density $\rho^b$, are monotonically decreasing functions
of $\rho_0^b$.}
\label{fig:2}
\end{figure}

We consider cases where the annealed fluid components are
characterised by the pair potential parameters
$\epsilon_{11}=\epsilon_{22}=2k_BT$, $\epsilon_{12}=1.8877k_BT$,
$R_{22}=0.665R_{11}$; this implies from Eq.\
(\ref{eq:pair_pot_mix_rule_1}) that $R_{12}=0.849R_{11}$. This
particular binary mixture exhibits bulk fluid--fluid phase separation and
was investigated in a number of previous studies
\cite{archer01,archer02,archer02el,archer03,archer05}.  The bulk phase
diagram is displayed in Fig.\ \ref{fig:1}. The parameters chosen for
the pair potentials characterising matrix particles are
$\epsilon_{\q\q}=2k_BT$, $\epsilon_{\q1}=k_BT$,
$\epsilon_{\q2}=0.8k_BT$, $R_{\q\q}=7R_{11}$; from Eq.\
(\ref{eq:pair_pot_mix_rule_1}) it follows that $R_{\q1}=5R_{11}$ and
$R_{\q2}=4.97R_{11}$. The reason for this choice of parameters
is that much is
known about the behaviour of an {\em annealed} fluid of such (big)
particles dissolved in the binary (solvent) fluid
\cite{archer02el,archer03,archer05}; hence we can make connections with
that body of work.
In Fig.\ \ref{fig:1} b) we display a portion of the phase
diagram, together with the loci in the {\em reservoir} phase diagram
of the state points which correspond [via Eq.\ (\ref{eq:mu_eff})] to
points on the binodal/spinodal for the fluid adsorbed in the matrix.
We display results for three densities of the quenched
particles, $\rho_\q^b R_{11}^3=0.0001$, 0.001 and 0.002 (or,
equivalently, $\rho_\q^b R_{\q\q}^3=0.0343$, 0.343 and 0.686). As the
matrix density $\rho_\q^b$ is increased, the binodal for the fluid
adsorbed in the matrix shifts to larger $x$, away from the reservoir
binodal. Note that one would obtain the
same shift for a matrix with a different set of pair potential
parameters and matrix densities, so long as the products $\rho_\q^b
\hat{v}_{\q n}(0)$ remain the same (recall Eq.\ (\ref{eq:mu_eff})).
For the above set of matrix pair potential parameters we
know \cite{archer02el,archer03,archer05} a {\em single} big particle
can be surrounded by a thick adsorbed film of the coexisting
phase. Furthermore, on decreasing $x$, for certain fixed total fluid
densities, there
can be a transition from a thin to a thick adsorbed film, 
see Fig.\ \ref{fig:1}. Given this information, and the
typical size of the (thick) adsorbed film, one would expect the matrix
condensation line (e.g.\ those in Fig.\ \ref{fig:1} b)) to be located
in the vicinity of the single particle thin--thick adsorbed film
transition line, provided the density of the quenched matrix particles
is sufficiently high that their adsorbed films (start to) overlap. Near
bulk coexistence, the
adsorbed film around a single big particle has a radius $\sim
7R_{11}$ \cite{archer02el,archer03,archer05}.  One would therefore expect
the matrix condensation line to be near the single particle thin--thick
adsorbed film transition line when the matrix density $\rho_\q^b
\simeq (4 \pi (7R_{11})^3/3)^{-1}= 0.7 \times 10^{-3} R_{11}^{-3}$. This
is indeed what we observe.  In fact, we find that the {\em spinodal} line
for the single particle thin--thick transition line lies very close to
the spinodal line (in the reservoir phase diagram) for a matrix of
density $\rho_\q^b = 10^{-3} R_{11}^{-3}$ -- see Fig.\ \ref{fig:2}b.
This observation provides physical insight into the mechanism driving
the condensation in the matrix:
condensation occurs when the big matrix particles are sufficiently close
that the adsorbed films around each of them can overlap.  This
means that condensation in the present soft core system is somewhat
different from the condensation that we alluded to earlier where the
condensation occurs in the gaps in the matrix -- i.e.\ as a surface
effect, where the films on (opposite) matrix surfaces join.  Rather, for the
present soft--core system, the onset of condensation is related to
the density profile of of the solvent around and inside a {\em single}
matrix particle.  One can therefore determine much about the fluid
behaviour in the bulk matrix from knowing the behaviour around a single
matrix particle.

In Fig.\ \ref{fig:2} we display the fluid densities, $\rho_n^b$,
$n=1,2$, inside a matrix of quenched particles with uniform density
for the case of a bulk fluid reservoir with total density
$\rho^b R_{11}^3=8.0$ and concentration $x=0.96$. Results are plotted
for a range of matrix densities $\rho_\q^b$. For
$\rho_\q^b R_{11}^3>2.3\times 10^{-3}$, we find
condensation of the coexisting fluid phase in the matrix. We also find
that $\rho_1^b$ (dashed--dotted line) is a non--monotonic function of
$\rho_\q^b$, whereas $\rho_2^b$ (dashed line) and the total density
$\rho^b=\rho_1^b+\rho_2^b$ (solid line), are
monotonically decreasing functions of $\rho_\q^b$.
For $\rho_0^b R_{11}^3 \gtrsim 0.1$ the fluid densities are very small;
both species are expelled from the matrix.
This behaviour is typical of that
found for reservoir state points lying on the right hand side of the
bulk binodal in Fig.\ \ref{fig:1}.

\section{Inhomogeneous fluid density profiles: planar interface}
\label{sec:matrix_surface}

In this section we determine the inhomogeneous fluid density profiles of
the binary fluid for a simple model of an interface in which the
quenched matrix particles are described by an average
density distribution:
\begin{equation}
\rho_\q(z) = 
\begin{cases}
\rho_\q^b \hspace{8mm} z \leq 0 \\
0 \hspace{10mm} z > 0,
\end{cases}
\label{eq:matrix_profile}
\end{equation}
i.e.\ the matrix particles are confined to the half--space with
(Cartesian) coordinate $z\leq 0$ and the fluid reservoir is at $z>0$.
The planar interface is located at $z=0$, where $z$ is the coordinate
perpendicular to the interface. The step function in Eq.\
(\ref{eq:matrix_profile}) is not the most realistic model for the matrix
density profile; one would expect some structure in the profile for $z
\rightarrow 0^{-}$ that will depend on how the quenched matrix is
formed. However, the simple choice (\ref{eq:matrix_profile})
allows us to explore some of the interfacial phenomena that can be
exhibited by the present
system. We do not expect our results to be changed qualitatively were
we to chose a more realistic profile.

\begin{figure}
\includegraphics[width=8cm]{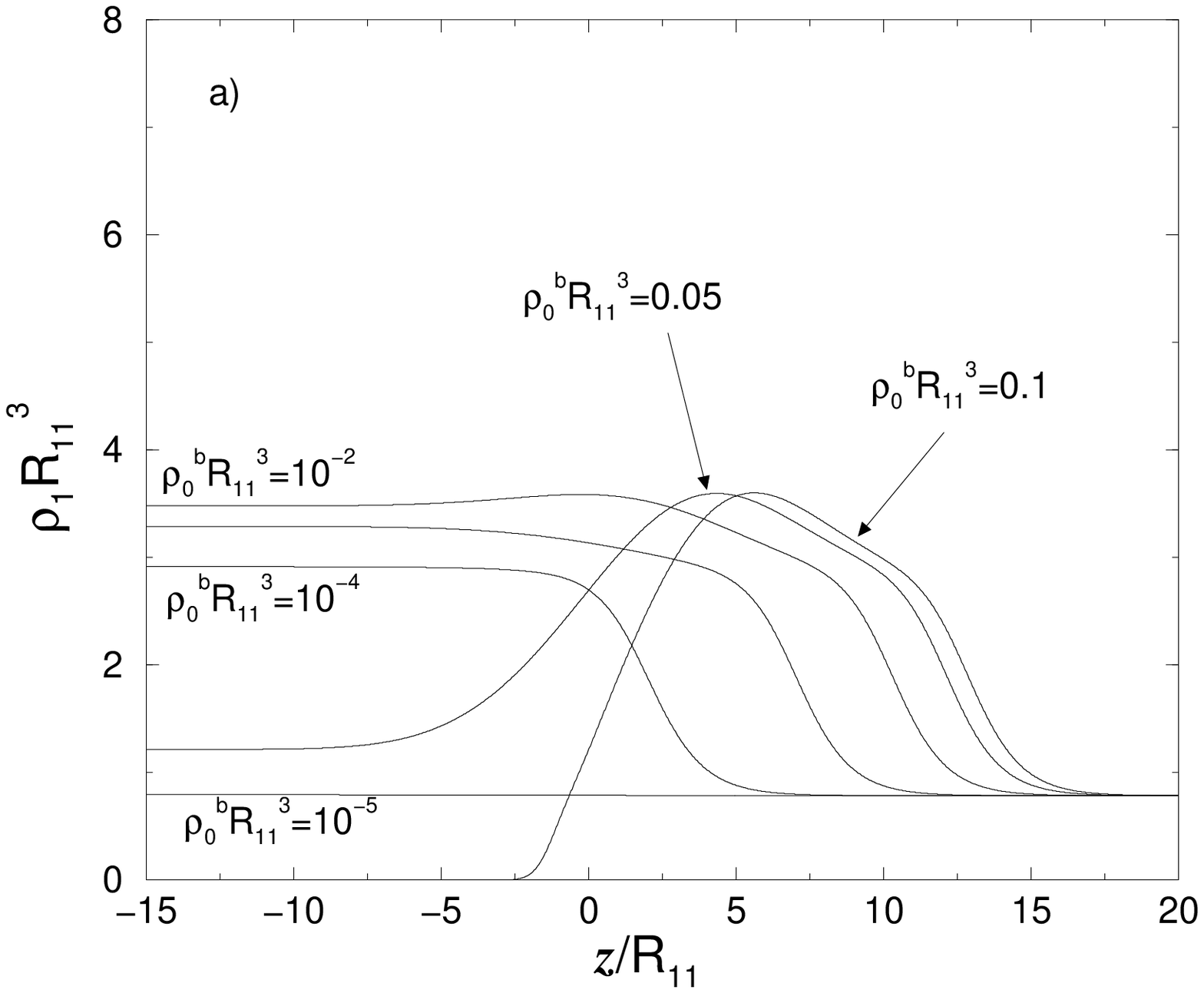}
\includegraphics[width=8cm]{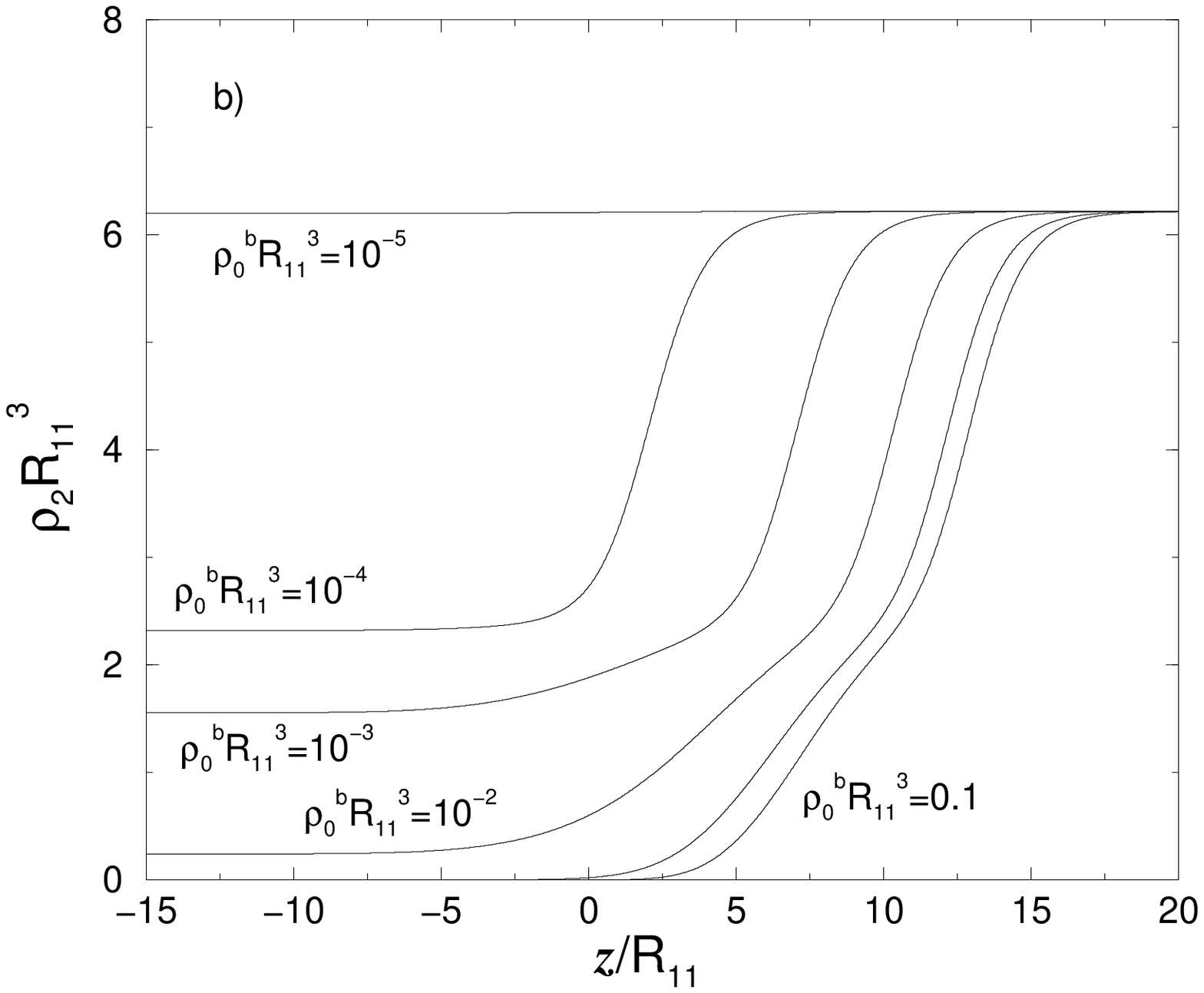}
\caption{Density profiles of the binary GCM
fluid at the planar interface between a
matrix with density profile given by Eq.\ (\ref{eq:matrix_profile}) and a
reservoir of fluid (at $z>0$) with fixed total bulk density
$\rho^b R_{11}^3=7.0$ and concentration $x=0.888$. The density profiles
for species 1 are shown in a), and those for species 2 in b).
The density profiles are calculated for matrix bulk densities
$\rho_\q^b R_{11}^3=10^{-5}$,
$10^{-4}$, $10^{-3}$, $10^{-2}$, 0.05 and 0.1. Note that
for intermediate densities the coexisting fluid phase is
condensed in the matrix. At this reservoir state point, there is a
wetting film extending from the interface at $z=0$ into the bulk
reservoir fluid for the higher matrix densities.}
\label{fig:3}
\end{figure}

In Fig.\ \ref{fig:3} we display the density profiles $\rho_n(z)$,
$n=1,2$, of the binary GCM fluid for cases where the reservoir fluid
has a fixed bulk density $\rho^b R_{11}^3=7.0$ and concentration
$x=0.888$ that is near to coexistence (located at $x_{coex}=0.88544$).
The density profiles are calculated for six matrix densities
$\rho_\q^b R_{11}^3=10^{-5}$, $10^{-4}$, $10^{-3}$, $10^{-2}$, 0.05
and 0.1. Provided the matrix density is
sufficiently high the coexisting fluid condenses in the matrix.
For large values of the matrix density the fluid is expelled from the
bulk of the matrix -- observe the results for $\rho_0^bR_{11}^3=0.1$.
The densities of the fluid in the bulk of the matrix ($z
\rightarrow - \infty$) behave in a similar manner to
the results displayed in Fig.\ \ref{fig:2}. Strikingly,
in addition to condensation in the bulk of the matrix,
a wetting film of the coexisting phase develops on the surface
of the matrix. This film is present in spite of the fact that the fluid
density in the matrix is very low, owing to the high matrix density.
Similar wetting behaviour was found for the present GCM
fluid mixture at a planar wall, for several choices of (repulsive)
wall--fluid potentials \cite{archer02}.

Since the decay of the effective
matrix potential (\ref{eq:V_eff}) into the bulk of the reservoir fluid
is short ranged, the thickness $l$ of the wetting film of the coexisting
phase on the surface of the matrix (i.e.\ $l$ is the distance of the
wetting film interface from $z=0$), should increase as
$l \sim -\xi_w \ln(\delta x)$,
where $\delta x =x-x_{coex}$ is the difference in concentration
between the reservoir concentration $x$ and the concentration at bulk
coexistence $x_{coex}$. $\xi_w$ is the true
correlation length in the bulk {\em coexisting} phase which is
wetting the wall (matrix)--reservoir interface \cite{archer02}.

\begin{figure}
\includegraphics[width=8cm]{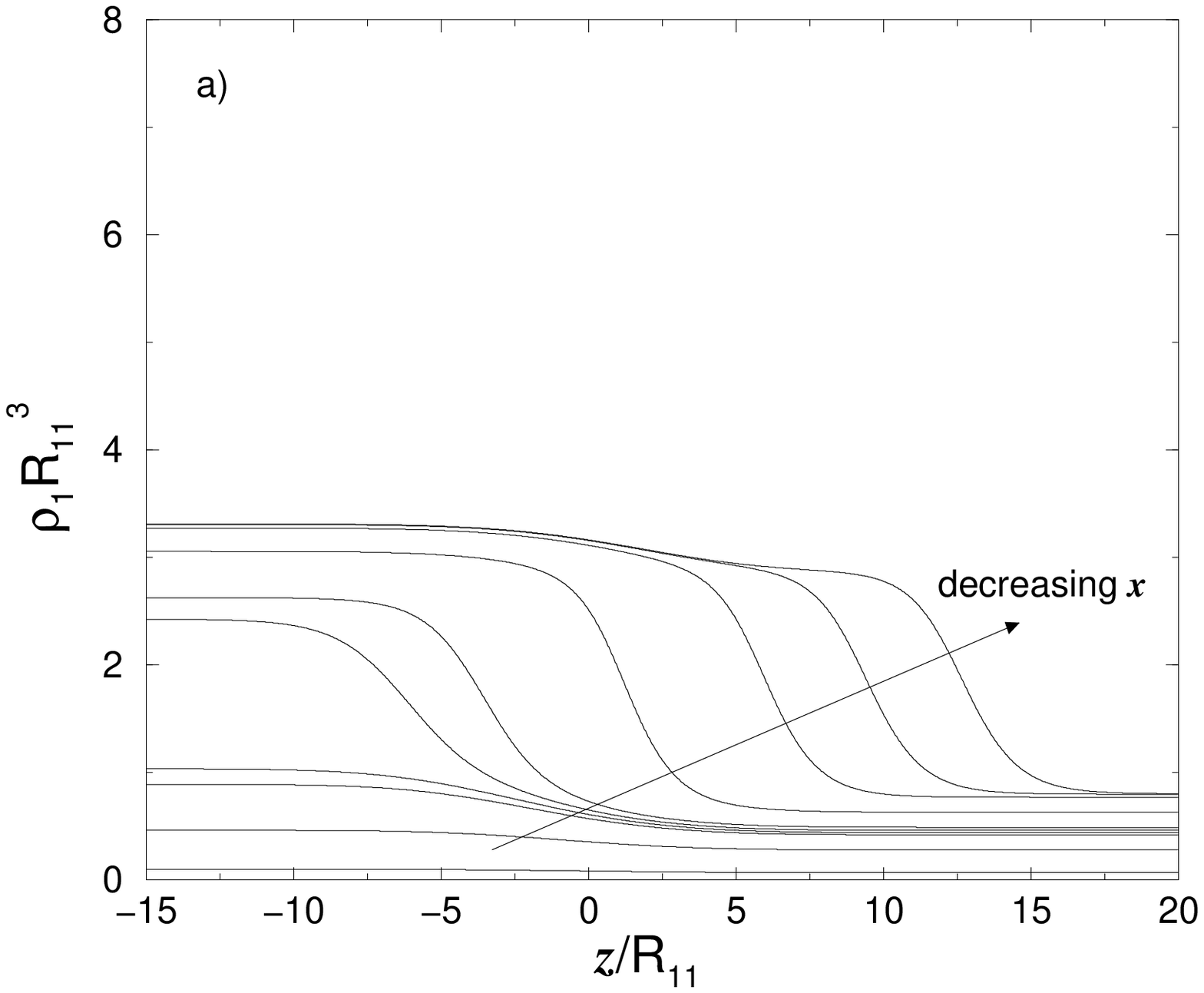}
\includegraphics[width=8cm]{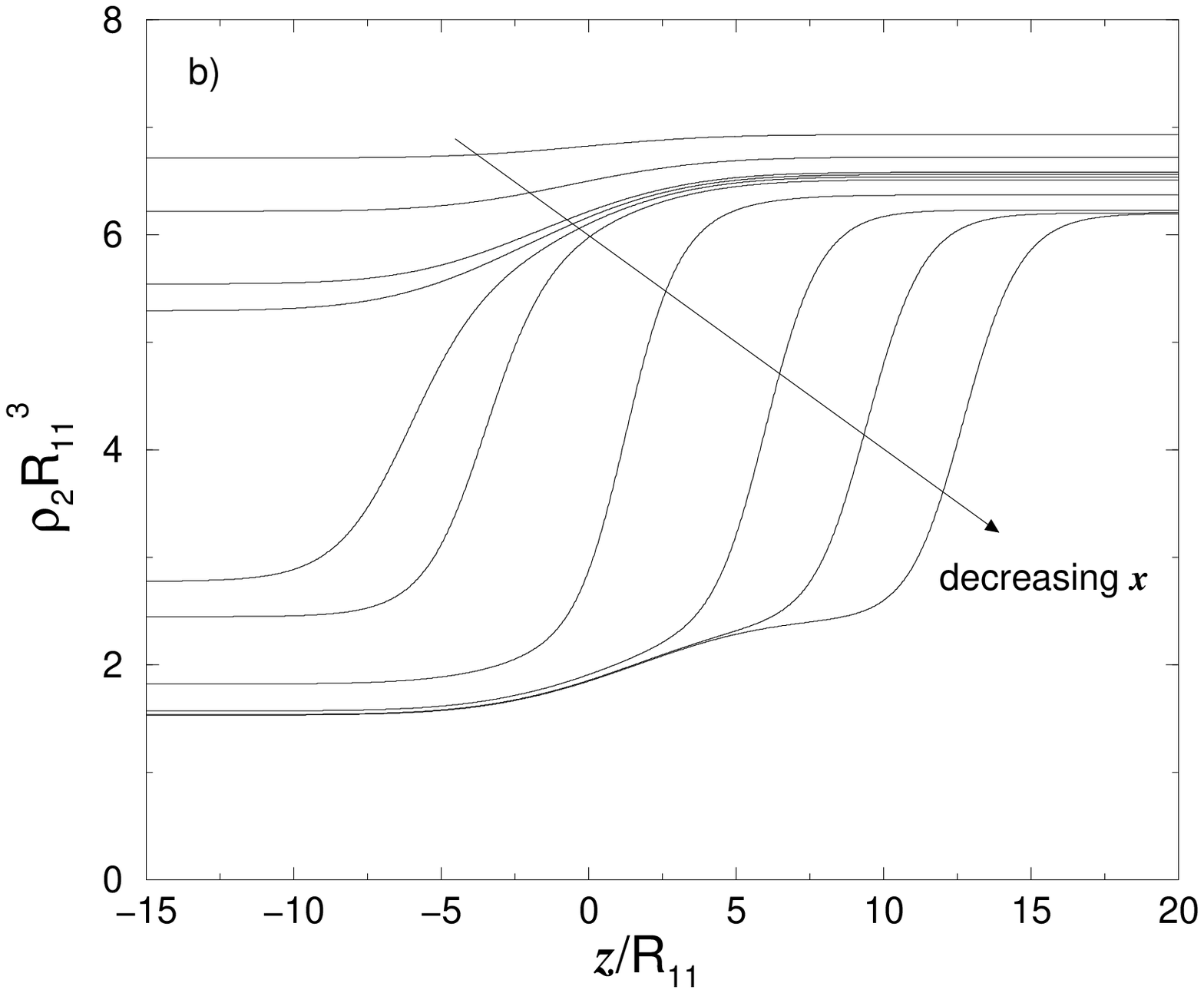}
\caption{Density profiles of the binary fluid at the planar interface between a
matrix with density profile given by Eq.\ (\ref{eq:matrix_profile}),
with $\rho_\q^b R_{11}^3=10^{-3}$. The bulk fluid reservoir
($z \rightarrow \infty$) has fixed total density $\rho^b R_{11}^3=7.0$.
The profiles are calculated for concentrations:
$x=0.99$, 0.96, 0.94, 0.937, 0.934, 0.93, 0.91, 0.89, 0.886 and 0.88545
(bulk coexistence is at $x_{coex}=0.88544$). These correspond to a series
of points
along path $A_r$ in Fig.\ \ref{fig:1} a). Note that between $x=0.937$
and 0.934, condensation of the coexisting phase occurs in the matrix.
Decreasing $x$ further, a wetting film develops on the surface of the
matrix.}
\label{fig:4}
\end{figure}

\begin{figure}
\includegraphics[width=8cm]{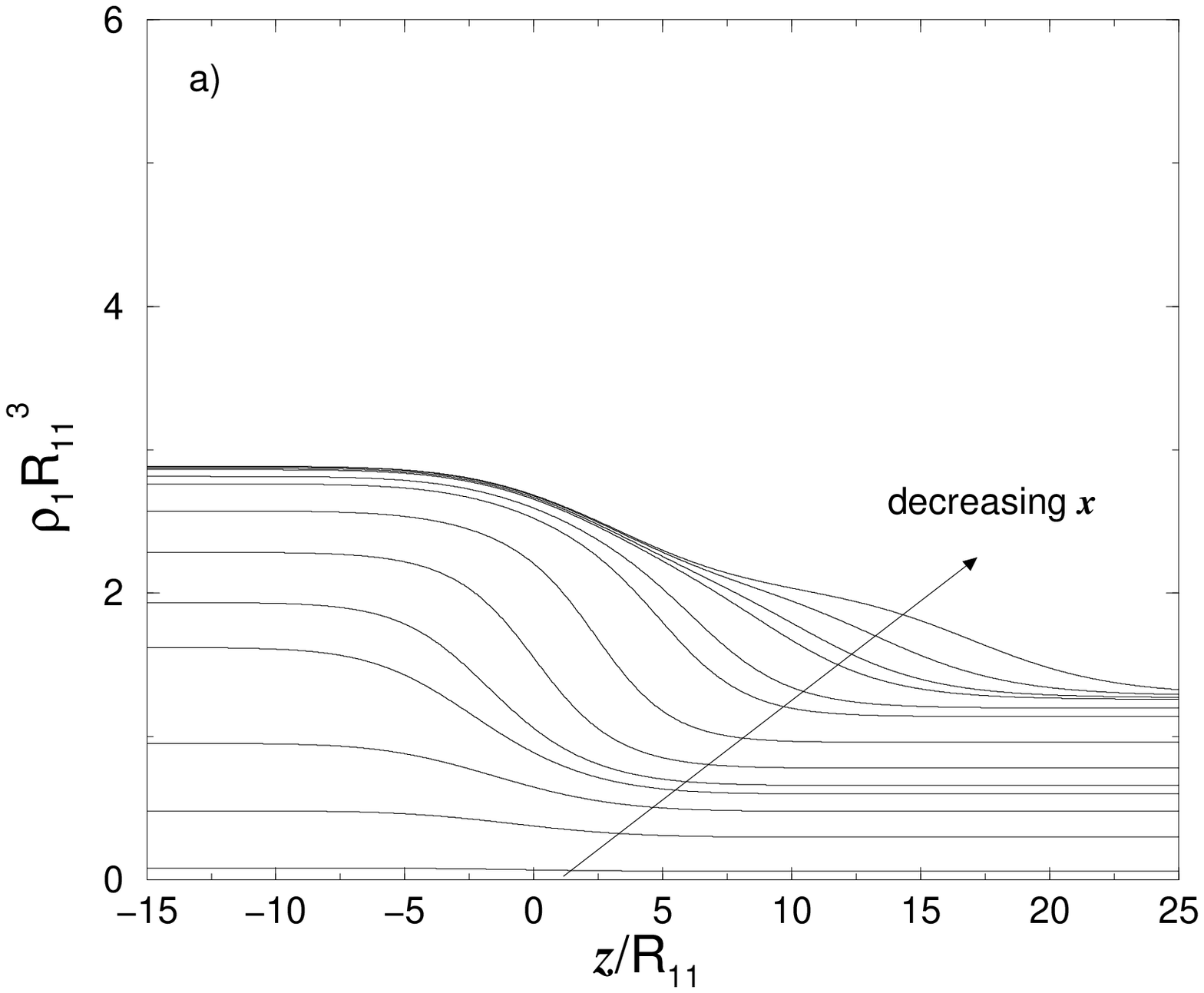}
\includegraphics[width=8cm]{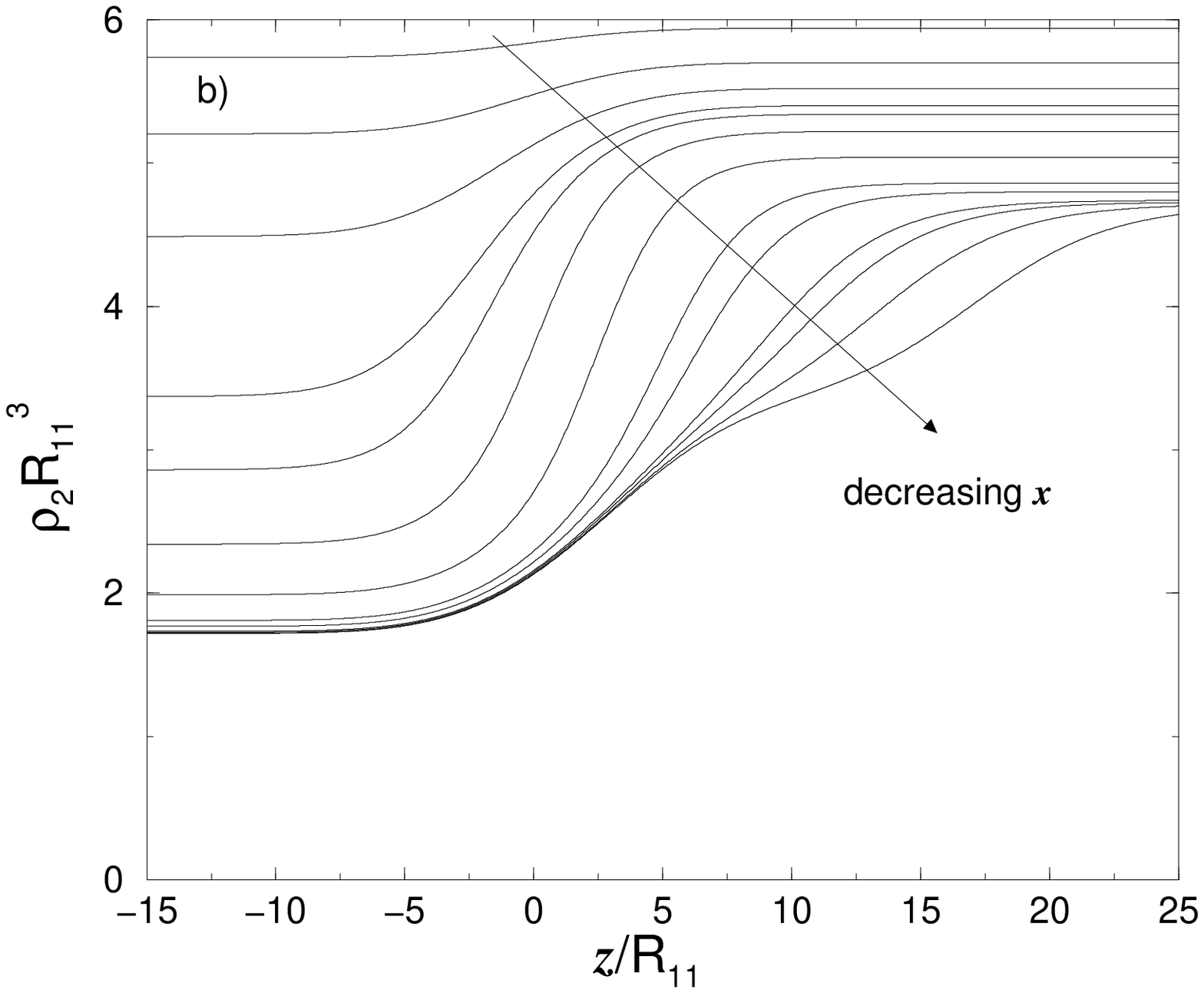}
\caption{Density profiles of the binary fluid at the planar interface
between a matrix with density profile given by Eq.\
(\ref{eq:matrix_profile}), with $\rho_\q^b R_{11}^3=10^{-3}$.
The bulk fluid reservoir ($z \rightarrow \infty$) has total density
$\rho^b R_{11}^3=6.0$. The profiles are calculated for
concentrations $x=0.99$, 0.95, 0.92, 0.9, 0.89, 0.87, 0.84,
0.81, 0.8, 0.79, 0.788, 0.786 and 0.785
(bulk coexistence is at $x_{coex}=0.7848$). These correspond to a series of
points along path $B_r$ in Fig.\ \ref{fig:1} a).
For decreasing $x$, a wetting film develops on the surface of the matrix. Note
that, contrary to the case in Fig.\ \ref{fig:4}, the density of the fluid
adsorbed in the matrix changes {\em continuously} with $x$.}
\label{fig:5}
\end{figure}

We investigate the wetting behaviour in more detail in Figs.\
\ref{fig:4} and \ref{fig:5} where we display the density profiles
of both species for the binary fluid at the planar
interface given by Eq.\ (\ref{eq:matrix_profile}) with fixed
$\rho_\q^bR_{11}^3=10^{-3}$. Results are presented
for reservoir state points corresponding to points along
paths $A_r$ and $B_r$ in Fig.\ \ref{fig:1} a). In Fig.\
\ref{fig:4} the profiles are calculated along path $A_r$ ($\rho^b
R_{11}^3=7$) for a range of concentrations approaching the reservoir
binodal. Between $x=0.937$ and 0.934 the coexisting phase condenses in
the matrix -- note the jump in the profiles.  As $x$ is decreased further,
a wetting film develops on the surface of the matrix. This increases in
extent as $x$ approaches its value at coexistence. In
Fig.\ \ref{fig:5} we display the density profiles for the same matrix
density but now for reservoir total density $\rho^b R_{11}^3=6.0$,
corresponding to path $B_r$ in Fig.\ \ref{fig:1} a).  Once again the
profiles are calculated for a range of concentrations approaching the
reservoir binodal and we observe a wetting film
developing on the surface of the matrix. Note the development of a
shoulder in the profile for $x=0.785$, close to coexistence.
Contrary to the case in Fig.\ \ref{fig:4}, the density of the fluid
adsorbed in the matrix changes {\em continuously} as $x$ is varied; there
is no jump. In both cases the wetting film thickens as
$x \rightarrow x_{coex}$, and we find that the film thickness
$l \sim -\xi_w \ln(\delta x)$.  In general, when the coexisting
phase is condensed in the matrix {\em and} the reservoir state point
is near to coexistence we expect the matrix--reservoir interface to be
wet completely by a film of the coexisting phase as $x \rightarrow
x_{coex}$.

\section{Conclusions}
\label{sec:conclusions}

We have considered the properties of a ternary GCM mixture, one of whose
species is quenched, modelling a rigid
matrix of (crosslinked) polymers. The two annealed species then
model a binary polymer solution that is immersed in the matrix.
Using a DFT treatment we have calculated both the behaviour of the binary
mixture in the bulk matrix and its properties at a planar
interface, described by the profile (\ref{eq:matrix_profile}), between
the matrix and a reservoir of the fluid.

In the first case where the fluid is adsorbed in the bulk matrix with
uniform density $\rho_\q^b$, we find that in the total density versus
concentration $(x,\rho^b)$ representation of the adsorbed fluid phase
diagram, the location of the fluid--fluid binodal is
independent of $\rho_\q^b$. This somewhat surprising result is a
direct consequence of the soft core nature of both the adsorbed fluid and
matrix particles as described within a second--order (in densities) free
energy -- see Eq.\ (\ref{eq:f}). In particular the contribution due to
the matrix is independent of the density of the annealed fluid. For given
fluid densities in the reservoir, the calculation of the fluid densities
in the matrix is very straightforward, and mapping out lines of (capillary)
condensation in the matrix, in the $(x,\rho^b)$ phase diagram, is readily
performed.

Applying a Helmholtz free
energy functional more accurate than the simple RPA functional
(\ref{eq:F_ex}), we would expect the fluid phase diagram in the
$(x,\rho^b)$ plane to depend explicitly on the matrix density
$\rho_0^b$. However, bearing in mind the accuracy of (\ref{eq:F_ex})
at high fluid densities
\cite{likos01II,lang00,louis00II,archer01,archer04,patrykiejew04}, we
expect only a weak dependence on $\rho_0^b$ in the true phase diagram of
the binary GCM immersed in a GCM matrix.

In earlier work \cite{archer02el,archer03,archer05}, we found that due to the
soft--core nature of the particles, condensation of the coexisting phase occurs
both on the surface
and {\em inside} a single big particle. Making connection with this
work, we found that for matrices composed of big GCM particles
with density $\rho_0 R_{00}^3 \sim 1$ (i.e.\ densities near the overlap
concentration), the mechanism for capillary condensation of the coexisting
fluid phase in the matrix is the joining of the thick films of the
coexisting phase that form around individual big GCM particles as bulk
coexistence is approached
\cite{archer02el,archer03,archer05}. This mechanism is somewhat different from
what occurs in hard core systems, where the wetting films grow on
the surface of big matrix particles and condensation occurs throughout
the matrix by the joining of these surface films.

Much of the simplicity inherent in the present theory stems from the
fact that the effect of the matrix on the fluid can be described in terms
of a shift in the effective chemical potential [c.f.\ Eq.\ (\ref{eq:mu_eff})]
and an effective one body external potential -- see Eq.\
(\ref{eq:V_eff}). Note that this simplification is not clear at the outset: a
particular configuration of matrix particles can, of course, be
treated as an external potential for the adsorbed fluid, but having averaged
over the
ensemble of possible matrix configurations, the description of the
matrix then enters on the level of the (ensemble average) one body
density profile of the matrix particles. That this problem can be subsequently
mapped to an effective one body potential, which is much simpler to
treat than the potential exerted by a particular configuration of
matrix particles, makes determining properties of the inhomogeneous
fluid in contact with an inhomogeneous matrix tractable. Following this
procedure we were able to calculate adsorption and wetting behaviour of the
binary mixture on the
surface of the matrix in contact with a reservoir of the fluid.
In the case of a reservoir at state points near to
coexistence, we found a rich interplay between the
fluid condensation in the bulk of the matrix and wetting
behaviour at the interface.
We conclude that the present DFT approach constitutes a
powerful approach for treating fluids in inhomogeneous random media.
Although the GCM model is a very simple one, it is physically realistic and the
results obtained from analysing its properties should provide much insight into
bulk and interfacial phenomena of fluids adsorbed in random porous media.

\section*{Acknowledgements}
A.J.A. acknowledges the support of EPSRC under grant number GR/S28631/01.


\end{document}